\documentclass[
preprint,
 amsmath,amssymb,
 aps,
floatfix
]{revtex4-2}
\usepackage[hidelinks,colorlinks=true,linkcolor=blue,citecolor=blue,urlcolor=blue]{hyperref}
\usepackage{graphicx}
\usepackage{dcolumn}
\usepackage{amssymb}
\usepackage{bm}
\usepackage{subfigure}
\usepackage{verbatim}
\usepackage{float}

\RequirePackage[normalem]{ulem}
\RequirePackage{color}

\begin{document}

\preprint{APS/123-QED}

\title{Multifractal Magnetoconductance Fluctuations in Mesoscopic Systems}
\renewcommand{\andname}{\ignorespaces}

\author{N. L. Pessoa,$^{1,2}$\ \textrm{A. L. R. Barbosa,}$^{3}$\ \textrm{G. L. Vasconcelos}$^{4}$\ and \textrm{A. M. S. Macêdo}$^{1}$}
 \affiliation{%
 $^{1}$Departamento\ de\ Física,\ Universidade\ Federal\ de\ Pernambuco,\textit{50670-901}\ Recife,\ Pernambuco,\ Brazil
}
\affiliation{%
 $^{2}$Centro\ de\ Apoio\ à\ Pesquisa,\ Universidade\ Federal\ Rural\ de\ Pernambuco,\ \textit{52171-900}\ Recife,\ Pernambuco,\ Brazil
}
\affiliation{%
 $^{3}$Departamento\ de\ Física,\ Universidade\ Federal\ Rural\ de\ Pernambuco,\ \textit{52171-900}\ Recife,\ Pernambuco,\ Brazil
}
\affiliation{%
 $^{4}$Departamento\ de\ Física,\ Universidade\ Federal\ do\ Paran\'a,\ \textit{81531-980}\ Curitiba,\ 
 Paran\'a,\ Brazil
}


\date{\today}

\begin{abstract}
We perform a multifractal detrended fluctuation analysis of the magnetoconductance data of two standard types of mesoscopic systems: a disordered nanowire and a ballistic chaotic billiard, with two different lattice structures. We observe in all cases that multifractality is generally present and that it becomes stronger in the quantum regime of conduction, i.e., when the number of open scattering channels is small. We argue that this behavior originates from correlations induced by the magnetic field, which can be characterized through the distribution of conductance increments in the corresponding ``stochastic time series," with the magnetic field playing the role of a fictitious time. More specifically, we show that the distributions of conductance increments are well fitted by $q$ Gaussians and that the value of the parameter $q$ is a useful quantitative measure of multifractality in magnetoconductance fluctuations.

\end{abstract}

\maketitle


\section{Introduction}

Universal conductance fluctuations (UCF) are among the most remarkable phenomena of mesoscopic physics \cite{RevModPhys.69.731,RevModPhys.89.015005}.
From a physical perspective, UCF can be traced back to quantum interference effects caused by multiple wave scatterings inside the sample.
Furthermore, UCF  are ubiquitous in that they have been observed in a great deal of  phase-coherent electron transport systems, from diffusive nanowires \cite{gold,PhysRevB.98.155407,PhysRevB.102.041107,PhysRevB.102.115105,Brazhkin_2020} to ballistic chaotic billiards \cite{sachrajda,taylor1,taylor2,taylor3,halfstadium,halfstadium2,HUANG20181}. 
There is,  however, a fundamental difference in the scattering mechanism of diffusive nanowires and ballistic chaotic billiards. In the former, the leading mechanism is impurity scattering, which induces Anderson localization when the sample's length exceeds the localization length, while in the latter one has elastic scattering taking place at the boundaries of the billiard. Notwithstanding this difference, UCF in both systems can be studied within a single perspective, by seeking to characterize more fully the statistical nature of the conductance fluctuations in all cases. 

UCF have been observed in a variety of experimental situations, such as by varying the strength of an externally applied magnetic field to a metallic sample, in which case they can also be called universal magnetoconductance fluctuations \cite{washburn}. Furthermore, they have also been observed by varying the Fermi energy of different types of samples and the numerical value of the variance of the conductance, measured in units of $e^2/h$, agrees with that calculated from magnetoconductance data. This means that UCF depend neither on the dimensions nor on the degree of disorder of the device, at least in a certain range of values of these parameters, hence the name universal \cite{UCF_Lee_Stone}. As a matter of fact, UCF can even be seen as sample-to-sample fluctuations in the conductance of samples with different disorder or border configurations \cite{UCF_Lee_Stone_Fukuyama}.

Among the many interesting features of UCF, one that we would like to point out is the fact that they have been shown to be fractals. 
For instance, it was reported in Ref.~\cite{gold} that the UCF of quasiballistic gold nanowires induced by a varying magnetic field have a fractal nature.
This fractality was attributed to the existence of long-lived states with chaotic trajectories close to regular classical orbits, which are characteristic of systems with a corresponding classical phase space that is neither fully chaotic nor integrable.

In the case of ballistic chaotic billiards, fractal behavior of UCF as a function of an externally applied magnetic field was observed for a soft wall stadium \cite{sachrajda} and a Sinai billiard \cite{sachrajda,taylor1,taylor2,taylor3} in high-mobility semiconductor heterojunctions. 
Furthermore, using a semiclassical approach to compute the transmission amplitudes of a nanostructured system \cite{jalabert}, Ketzmerick \cite{ketzmerick} showed that the UCF of ballistic chaotic billiards are fractals and computed the fractal dimension $D_F$. More specifically, from the magnetoconductance, i.e., the conductance increments $\Delta G$ as a function of magnetic field variation $\Delta B$ at a fixed Fermi energy,
 \begin{equation}
 \Delta G = G(B+\Delta B) - G(B),\label{GB}
 \end{equation}
Ketzmerick proved that the second moment of the magnetoconductance scales with $\Delta B$ as
$ \langle (\Delta G)^2 \rangle \sim (\Delta B)^\gamma$, where $\gamma = 1 - D_F/2$.
This behavior was shown to be a consequence of the power-law tail of the probability density, $P(T) \sim T^{-\gamma}$, for an electron to stay in the cavity up to time $T$, which is a known feature of systems with a hierarchical phase space structure which have a mixed (regular and chaotic) dynamics.

Recently,  the fractal nature of UCF came up again after a somewhat unexpected experimental detection of multifractality in the conductance fluctuations of a single-layer graphene sample \cite{Amin2018}. 
In practical terms, multifractality  means that it is not enough to determine just a single scaling exponent, such as $\gamma$, which is related to the usual fractal dimension \cite{ketzmerick} (see above), to describe fluctuations, but instead an infinite number of scaling exponents becomes necessary to fully characterize the fluctuation statistics.
Differently from the experimental observations of fractality, as in Ref.~\cite{gold}, where the number  of  propagating  wave  modes  in  the  leads are large (semiclassical regime), the multifractal behavior emerges when there is just one propagating  wave  mode  in  the  leads (extreme quantum regime), more specifically, close to the charge-neutrality (Dirac) point of single-layer graphene  \cite{Amin2018}. 
Besides, the authors of Ref.~\cite{Amin2018} conjectured that the experimental observation of multifractality might be an evidence of an incipient Anderson localization near the Dirac point, since this appears to be the most plausible cause for the multifractal behavior. 
However, as we will show with the results of this paper, this conjecture deserves a deeper analysis. More specifically, we will argue that multifractality in UCF stems from certain quantum correlations induced by the magnetic field. 


In this work, we present a systematic numerical study of the multifractal spectrum of the conductance fluctuations obtained by varying a perpendicularly applied magnetic field to two kinds of mesoscopic graphene devices: disordered nanowires and ballistic chaotic billiards (see Fig.~\ref{sistemas_grafeno}). 
Furthermore, we also present results of standard confined two-dimensional electron gases (dots and wires), in order to make direct comparisons with the results for our graphene systems.
Our analysis suggests that multifractality is a common feature of conductance fluctuations of all types of mesoscopic devices in the quantum regime of conduction.
It means that magnetoconductance fluctuations of disordered nanowires and ballistic chaotic billiards show multifractal behavior when the number of open scattering channels is sufficiently small, which in turn led us to discard proximity to phase transitions as a necessary cause of multifractal conductance fluctuations (MCF).
Instead, we argue that MCF are related to certain types of quantum correlations \cite{PhysRevLett.93.014103, Novaes_2013}, which are present in the quantum regime and are lost when the system enters into the semiclassical regime of conduction. To be more specific, it has been shown using the trajectory-based semiclassical approach \cite{PhysRevLett.89.206801,PhysRevE.85.045201,Kuipers_2013} that characteristics of quantum transport observables can be calculated via sums over classical scattering trajectories. Furthermore, correlations between such trajectories can be organized diagrammatically and have been shown to disappear in the dominant semiclassical regime (large number of scattering channels) and give exact results in the extreme quantum regime (small number of scattering channels). We believe that these trajectory correlations induced by the magnetic field are the ones that are captured by the multifractal analysis and by the fat tails of the distribution of conductance increments, which incidently are one of the features of intermittency in fluid turbulence.

\section{Methods}\label{sec2}

In this section, we briefly introduce the scattering model used to obtain the magnetoconductance series as well as the method we used to perform the multifractal analysis of the numerical data.

\subsection{Scattering model}

\begin{figure}
   \centering
   \includegraphics[width=0.8\linewidth]{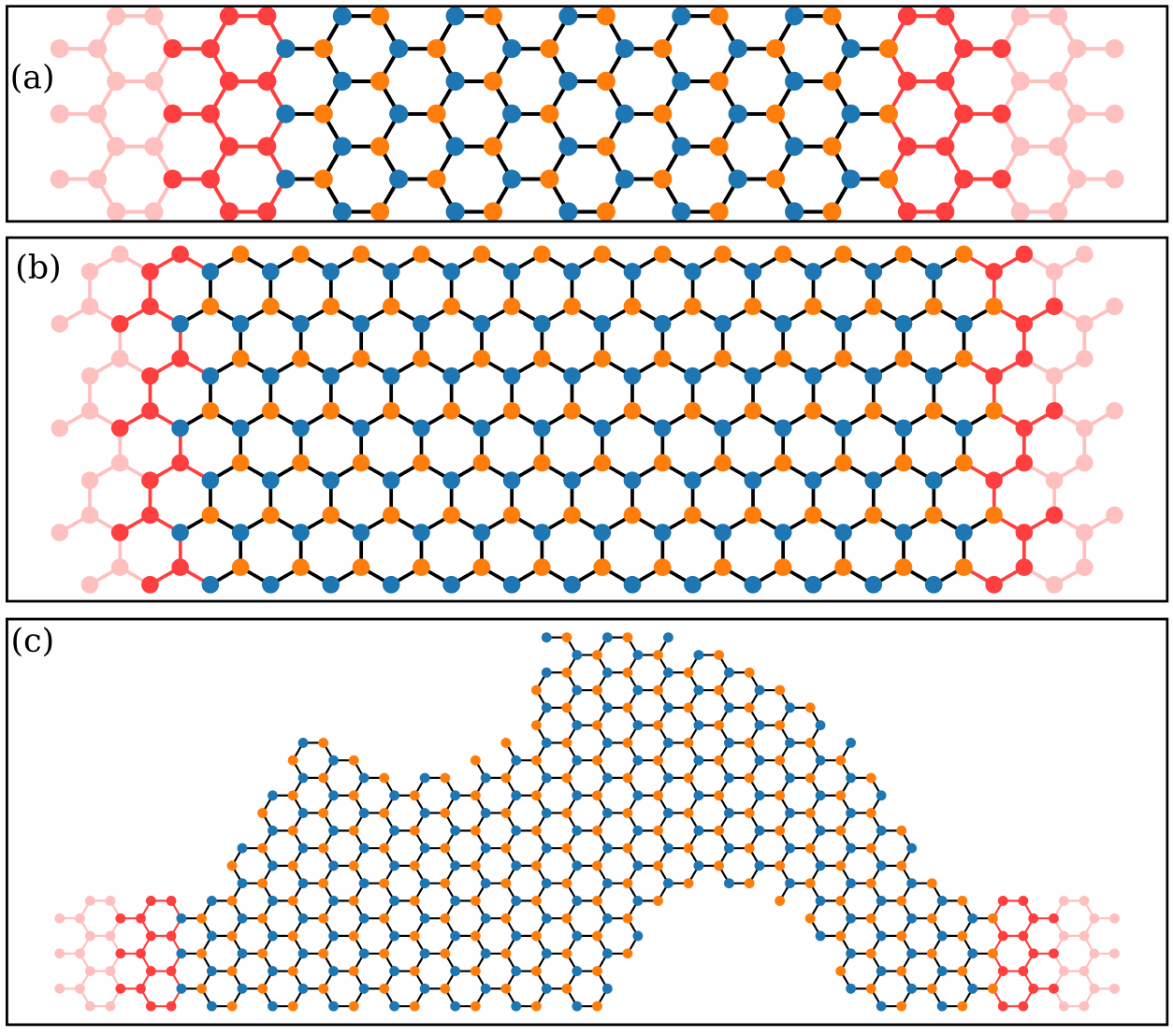}
    \caption{Schematics of disordered graphene nanowires connected to leads (red) with two boundary configurations: (a) armchair and (b) zigzag, and of (c) a graphene chaotic billiard.}
   \label{sistemas_grafeno}
\end{figure}

We consider electronic transport through three different graphene mesoscopic devices connected to two leads, as illustrated in Fig.~\ref{sistemas_grafeno}. Two of them are disordered graphene nanowires, namely (a) armchair (AGNR) and (b) zigzag nanoribbons (ZGNR), and the third is (c) a graphene chaotic billiard (GCB).
Electronic transport through the devices can be described by the scattering matrix 
\begin{equation}
    S= \left(
\begin{array}{cc}
\ r   & t'    \\
t  &r' 
\end{array}
\right),
\end{equation}
where $t$ ($t'$) and $r$ ($r'$) are transmission and reflection blocks, respectively. 
The conductance can be calculated from the Landauer-B\"{u}ttiker relation 
\begin{equation}
G = \frac{2e^2}{h}\textrm{Tr}(t^\dagger t),
\end{equation}
 which is valid in the linear regime and at low temperatures. The numerical calculations of the conductance were performed using the Kwant software \cite{kwant}, which is a Green's function---based algorithm that computes, among other quantities, the transport observables of a system described via a tight-binding approach. The tight-binding Hamiltonian for a single-layer graphene device is given by
\begin{equation}\label{hamilton_disc_graphene}
H = t_0 \sum_{\langle i,j \rangle}c_i^\dagger c_je^{i\theta_{ij}}+\sum_{i}\epsilon_{i} c_i^\dagger c_i,
\end{equation}
where the indices $i$ and $j$ run over all lattice sites and $\langle i,j \rangle$ denotes nearest neighbors. The first term of Eq.~(\ref{hamilton_disc_graphene}) represents the usual electron hopping between lattice sites, $c_i$ ($c_i^\dagger$) are the annihilation (creation) operators, $t_0$ is the hopping energy, which has a typical experimental value around 2.8 eV, and $\theta_{ij}=-(e/\hbar) \int_i^j \textbf{A}\cdot\textbf{dl}$ is the Peierls phase factor related to the externally applied magnetic field. We use the gauge condition given by $\textbf{A} = (-By,0,0)$, so that $\textbf{B}=(0,0,B)$ is the magnetic field, which we measure in terms of the dimensionless magnetic flux $\phi  =\Phi/\Phi_0$, where $\Phi=Ba^2$ is the flux through an enclosed area $a^2$ of the lattice and $\Phi_0 = h/e$ is the flux quantum.
The second term of Eq.~(\ref{hamilton_disc_graphene}) is an Anderson disorder term that is realized by an on-site electrostatic potential $\epsilon_{i}$ which varies randomly from site to site according to a uniform distribution in the interval  $(-U/2,U/2)$, where $U$ indicates the disorder strength.

\subsection{Multifractal analysis}

We shall first introduce a dimensionless conductance $g=G/(2e^2/h)$, which will be used throughout this work. Now let $g_k\equiv g(\phi_k)$, $k=1,...,M,$ be the conductance series obtained by varying the magnetic flux $\phi$, at fixed increments $\Delta\phi$, 
for a given system, where $M$ is the total number of values of $\phi$ chosen; see below for the specific increments in the magnetic flux for each system. As the magnetic field can be viewed as playing the role of a fictitious time, we may regard the dataset $\{g_k\}$ as an effective ``time series." Our treatment  of the data series is based on the multifractal detrended fluctuation analysis (MF-DFA) \cite{mfdfa}. In order to apply the MF-DFA, we first divide the time series into $N_s=M/s$ nonoverlapping windows of size $s$. The main idea of this method is then to determine the $p$th order fluctuation function
\begin{equation}\label{fq}
    F_p(s) = \left(\frac{1}{2N_s}\sum_{j=1}^{2N_s}\left[F^2_s(j)\right]^{p/2}\right)^{1/p},
\end{equation}
for an appropriate range of values of the exponent $p$, say from some finite negative value to its opposite (positive) value. In Eq.~(\ref{fq}), we define
\begin{equation}
    F^2_s(j) = \frac{1}{s}\sum_{i=1}^s[\tilde{g}((j-1)s+i)-P_j(i)]^2,
    \label{F2s}
\end{equation}
where
\begin{equation}
    \tilde{g}(i) = \sum_{k=1}^i(g_k-\langle g\rangle)
\end{equation}
represents a zero-mean profile  of the original series $g_k$ and $P_j(i)$ is a polynomial fit to the profile $\tilde{g}(i)$ over the $j$th segment of size $s$. In our analysis, we have considered only linear fits, so that Eq.~(\ref{F2s}) implements a local linear detrending in each segment.

Once we have determined the set of functions $F_{p}(s)$, we study their scaling with the window size $s$ according to the following relation:
\begin{equation}
    F_p(s) \sim s^{H(p)} \label{FH},
\end{equation}
where $H(p)$ is the generalized Hurst exponent. Here we have considered $p$ ranging from $-5$ to $5$ with steps of 0.2. We recall that $H(2)$ is the Hurst exponent defined in the standard fractal analysis \cite{Mandelbrot}. If $H(p)$ is $p$ dependent, we say that the corresponding series is multifractal, while if $H(p)$ does not change as $p$ is changed, we say that we have a monofractal series. We also define
\begin{equation}
    \tau(p) = pH(p)-1,
\end{equation}
such that we have the multifractal singularity spectrum $f(\alpha)$ defined as a Legendre transformation of $\tau(p)$:
\begin{equation}
    f(\alpha) = \alpha p-\tau(p). \label{falfa}
\end{equation}
From the point of view of the singularity spectrum $f(\alpha)$, we know that multifractal time series are characterized by a broad $f(\alpha)$, while monofractal ones by a narrow $f(\alpha)$. In other words, the strength of the multifractality can be seen as the width of $f(\alpha)$, $\Delta\alpha=\alpha_{\textrm{max}}-\alpha_{\textrm{min}}$, such that as $\Delta\alpha \rightarrow 0$, we have a loss of multifractality.

\section{Results}\label{sec_results}

In this section, we show the numerical calculation of the conductance series as a function of the magnetic flux $\phi$. We begin by analyzing the two types of disordered graphene nanowires, followed by the GCB (see Fig.~\ref{sistemas_grafeno}).

\subsection{Disordered nanowires}\label{sec3A}

The disordered graphene nanowires were designed as armchair (AGNR) and zigzag (ZGNR) nanoribbons, as illustrated by Fig.~\ref{sistemas_grafeno}. The length of the AGNR and ZGNR samples are  $L_A = 127 a$ and $L_Z = 124 \sqrt{3} a$, respectively, while their widths are  $W_A = 11 \sqrt{3}a$ and $W_Z = 67 a/2$, where $a = 2.46$ $\textrm{\AA}$ is the graphene lattice constant.
In order to implement the numerical calculations of the magnetoconductance series, the Fermi energy and disorder strength $U=1.30t_0$ are kept fixed, while the magnetic flux is varied with increments $\Delta\phi=1\times10^{-5}$.

\begin{figure*}
    \centering
    \includegraphics[width=0.95\linewidth]{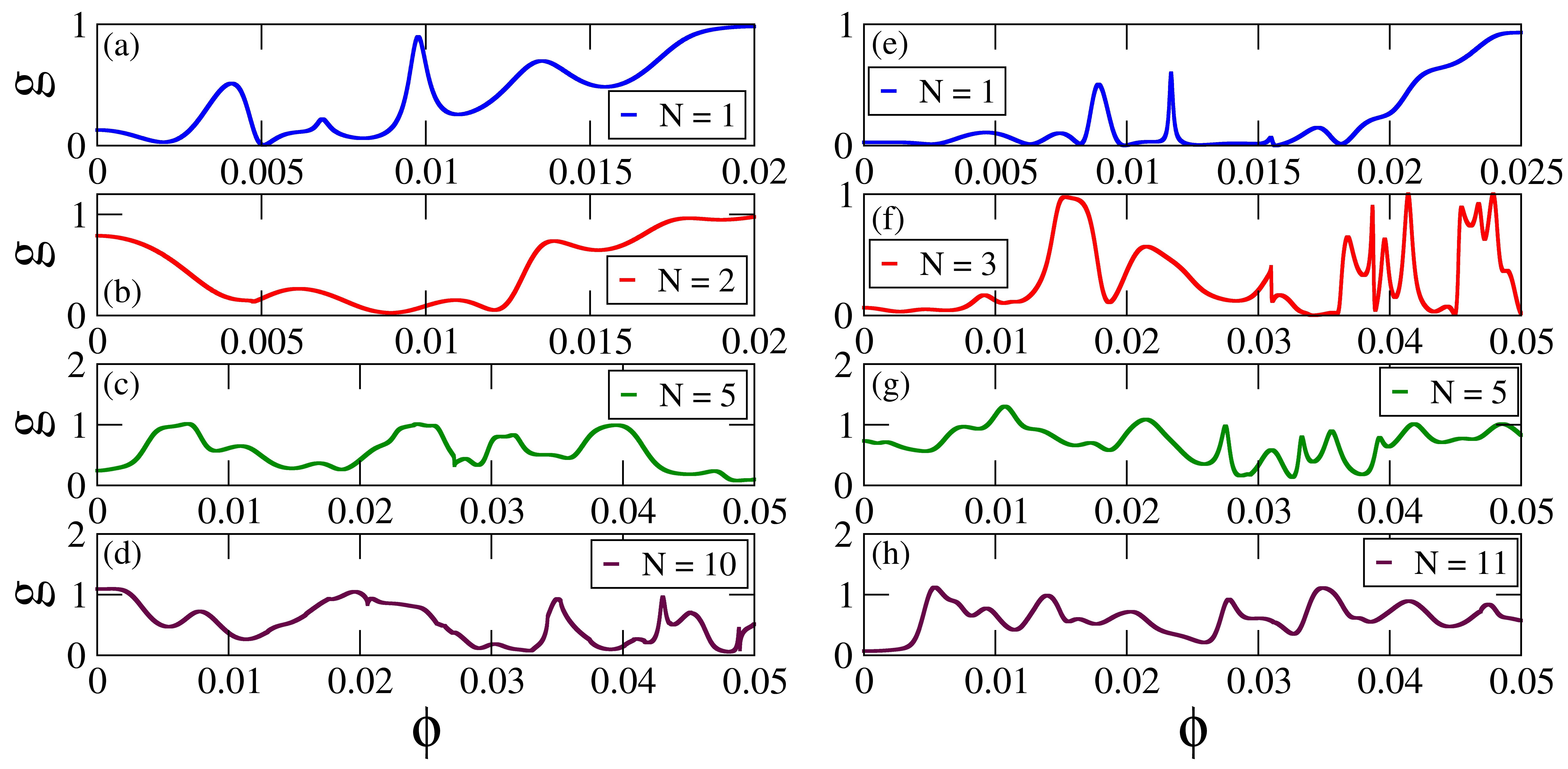}
    \caption{Single realizations of the magnetic field-induced conductance fluctuations of graphene disordered wires with armchair boundaries with (a) $N=1$, (b) $N=2$, (c) $N=5$, (d) $N=10$ propagating channels are shown on the left panel and with zigzag boundaries with (e) $N=1$, (f) $N=3$, (g) $N=5$, (h) $N=11$ propagating channels are shown on the right panel. The conductance is given in units of $2e^2/h$ and the magnetic flux in units of $Ba^2/(h/e)$.}
    \label{series_fios_grafeno}
\end{figure*}

Figure \ref{series_fios_grafeno} shows typical conductance series as functions of magnetic flux of AGNR (left) and ZGNR (right) for four different values of the number $N$ of propagating wave modes in the leads, which are $N=1,2,5,10$ for AGNR and $N=1,3,5,11$ for ZGNR. 
The integer number $N$ is defined as $N = k_F W/\pi$, where $W$ is the width of either lead and $k_F$ is the Fermi wave number, which implies that the Fermi energy is tuned for a specific value of $N$ to be achieved.

\begin{figure}[t]
    \centering
    \includegraphics[width=0.82\linewidth]{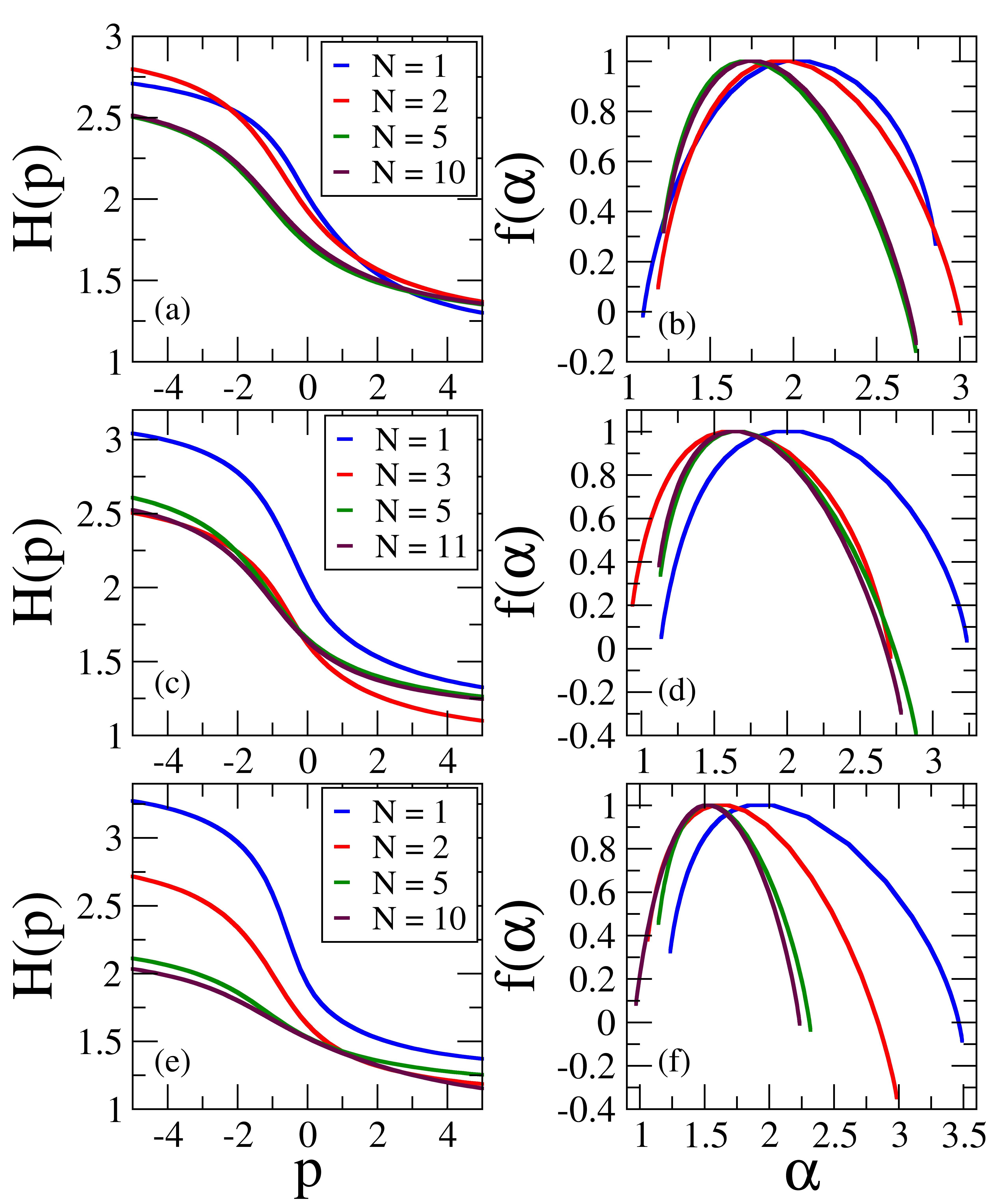}
    \caption{Left panel shows the mean generalized Hurst exponent $H(p)$ of the conductance fluctuations of (a) graphene disordered wires with armchair boundaries with $N=1,2,5,10$ propagating channels,  (c) graphene disordered wires with zigzag boundaries with $N=1,3,5,11$ propagating channels, and (e) two-dimensional electron gases with a square lattice with $N=1,2,5,10$ propagating channels. The right panel shows the multifractal singularity spectrum of the conductance fluctuations of (b) graphene disordered wires with armchair boundaries with $N=1,2,5,10$ propagating channels,  (d) graphene disordered wires with zigzag boundaries with $N=1,3,5,11$ propagating channels, and (f) two-dimensional electron gases with a square lattice with $N=1,2,5,10$ propagating channels.}
    \label{Hq_e_falfa_fios}
\end{figure}

Applying the multifractal analysis to 10 different disorder realizations of the conductance series, we obtained the results shown in Fig.~\ref{Hq_e_falfa_fios}. 
Figures \ref{Hq_e_falfa_fios}(a) and \ref{Hq_e_falfa_fios}(c) show the mean Hurst exponent, given by Eq.~(\ref{FH}), as a function of the order $p$ for AGNR and ZGNR, respectively. 
Both show a strong variation of $H(p)$ with $p$ at the extreme quantum regime ($N=1$). 
However, there is still a variation for larger values of $N$, but it becomes weaker as $N$ increases, i.e., when the system enters into the semiclassical regime. Hence the results suggest that conductance fluctuations are multifractal, and that this behavior is stronger in the quantum regime of conduction.

Furthermore, Figs.~\ref{Hq_e_falfa_fios}(b) and \ref{Hq_e_falfa_fios}(d) show the mean multifractal spectrum $f(\alpha)$, Eq.~(\ref{falfa}), as a function of the $\alpha$ parameter for AGNR and ZGNR, respectively. 
The narrowing of the singularity spectrum $f(\alpha)$, which can be measured by the decrease of $\Delta\alpha$, indicates the weakening of the multifractality of the magnetoconductance fluctuations as $N$ increases from $1$ to higher values, for both AGNR and ZGNR.
This means that, in the extreme quantum regime, the system exhibits MCF, i.e., conductance fluctuations have strong multifractal behavior, irrespective of any proximity to a localization transition. On the other hand, for $N>1$ the multifractality weakens and tends to become monofractal in the semiclassical regime ($N\gg 1)$. 

To finalize this section, we also report a multifractal analysis of the conductance series of laterally confined two-dimensional electron gases (2DEG) with a square lattice, with length $L=198a$ and width $W=30a$, and disorder strength $U=0.65t_0$, as in \cite{PhysRevB.98.155407}. 
The conductance series as a function of magnetic flux were obtained with increments given by $\Delta\phi=2\times10^{-5}$. Figures \ref{Hq_e_falfa_fios}(e) and \ref{Hq_e_falfa_fios}(f) show the mean Hurst exponent and multifractal spectrum of the 2DEG, respectively. 
These results suggest that the conductance fluctuations of 2DEG devices have multifractal behavior in the extreme quantum regime ($N=1$) and that this multifractality becomes weaker in the semiclassical regime, as shown above for graphene nanoribbons.
This indicates that MCF are not a specific feature of graphene nanowires.

\subsection{Chaotic Billiards}

\label{sec:GCB}

Now, we will discuss the multifractal analysis of the conductance series of ballistic graphene chaotic billiards (GCB), whose schematics are shown in Fig.~\ref{sistemas_grafeno}(c).
The tight-binding Hamiltonian is given by Eq.~(\ref{hamilton_disc_graphene}) without the Anderson disorder term, i.e, the disorder strength is kept null, $U=0$.
The GCB design follows that of Refs.  \cite{halfstadium,halfstadium2}.

\begin{figure}
    \centering
    \includegraphics[width=0.85\linewidth]{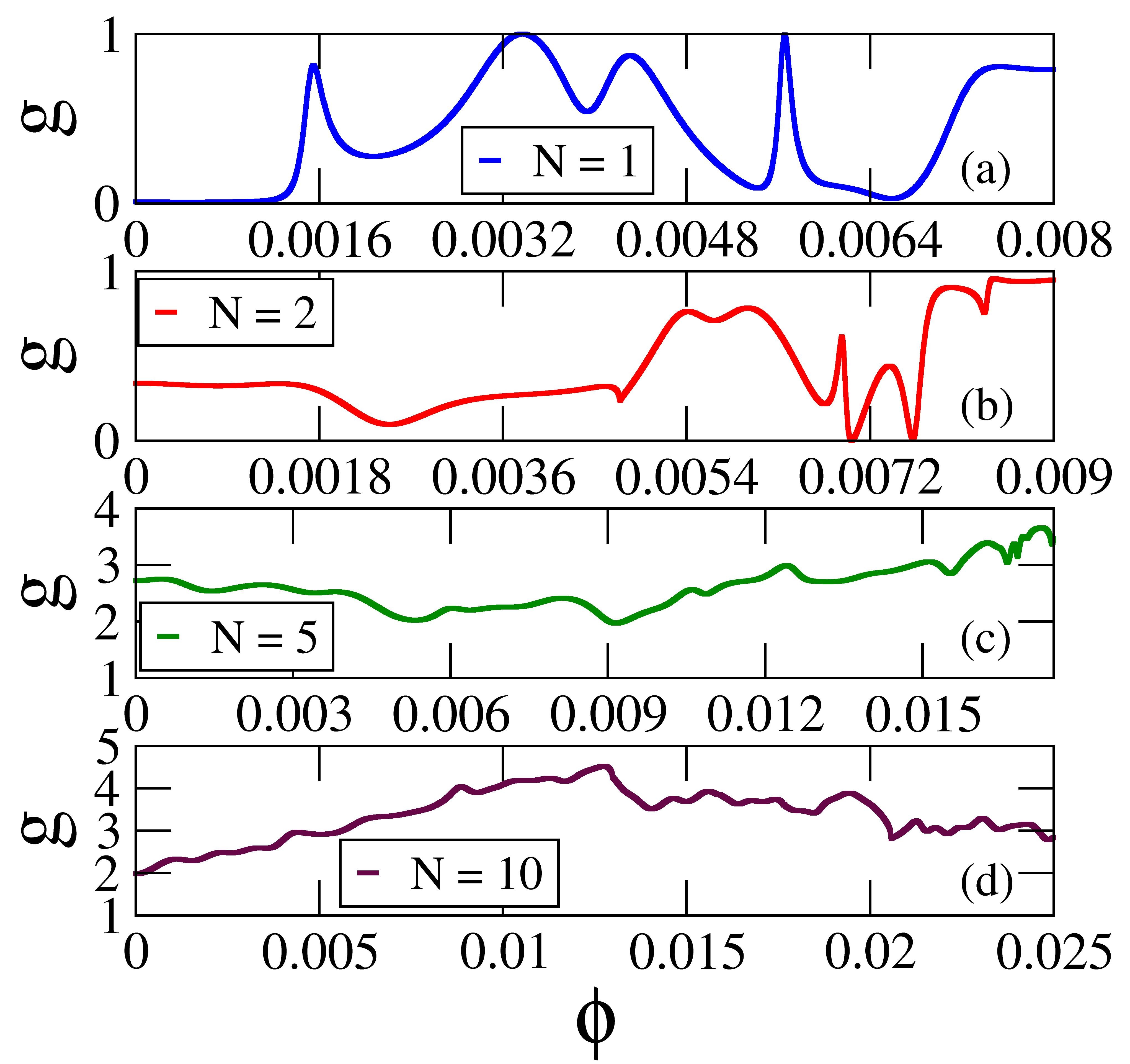}
    \caption{Single realizations of the magnetic field-induced conductance fluctuations of graphene chaotic billiards with (a) $N=1$, (b) $N=2$, (c) $N=5$, (d) $N=10$ propagating channels. The conductance is given in units of $2e^2/h$ and the magnetic flux in units of $Ba^2/(h/e)$.}
    \label{series_ponto_grafeno}
\end{figure}

\begin{figure}
    \centering
    \includegraphics[width=0.95\linewidth]{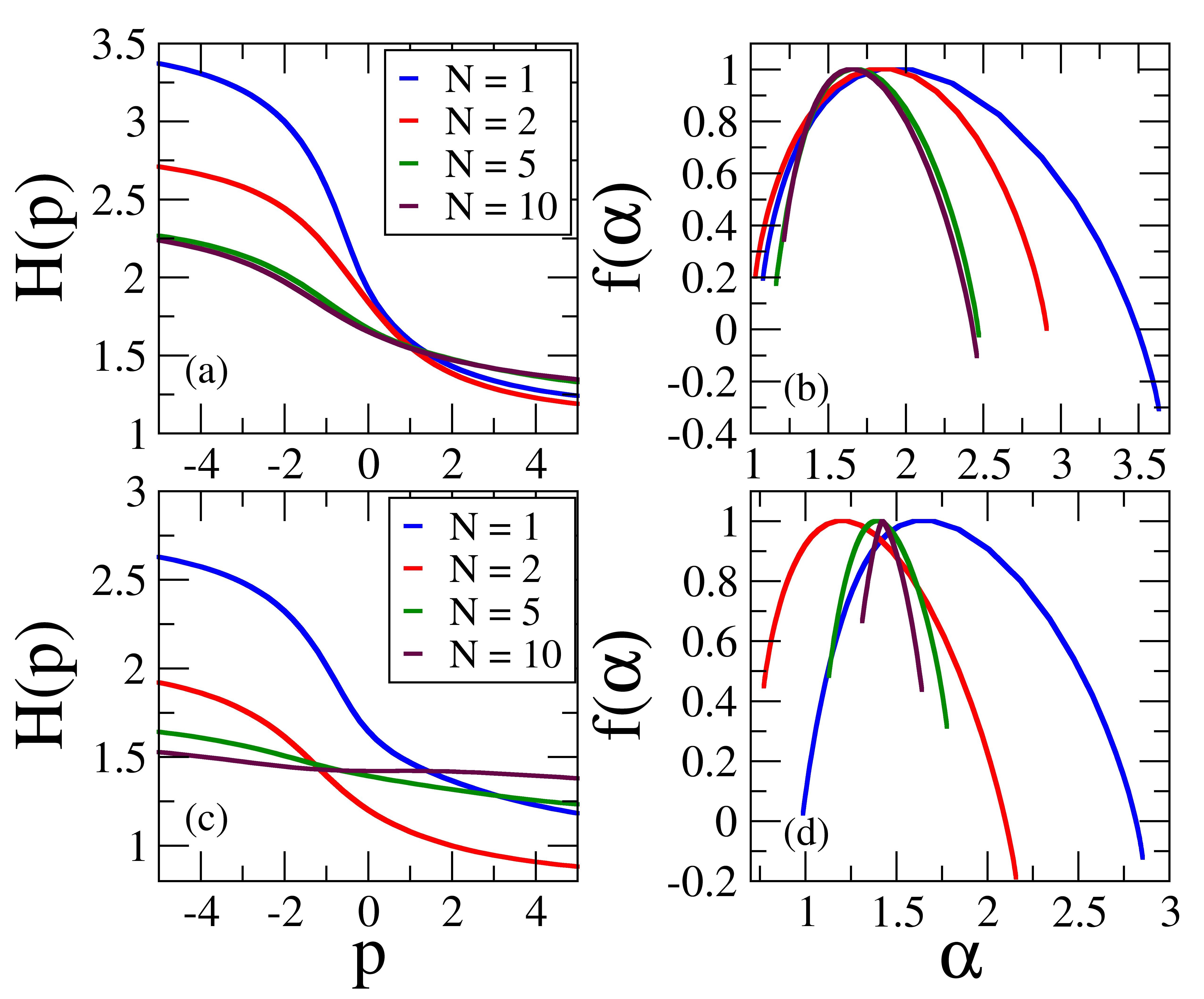}
    \caption{Left panel shows the mean generalized Hurst exponent $H(p)$ of conductance fluctuations of (a) graphene chaotic billiards with $N=1,2,5,10$ propagating channels and (c) chaotic billiards with a square lattice with $N=1,2,5,10$ propagating channels. The right panel shows the multifractal singularity spectrum of conductance fluctuations of (b) graphene chaotic billiards with $N=1,2,5,10$ propagating channels and (d) chaotic billiards with a square lattice with $N=1,2,5,10$ propagating channels.}
    \label{Hq_e_falfa_pontos}
\end{figure}

Figure \ref{series_ponto_grafeno} shows typical conductance series as functions of the magnetic flux of GCB. They were obtained by tuning the Fermi energy in order to set the number of propagating modes in the leads to $N=1,2,5,10$, and using magnetic flux increments given by $\Delta\phi = 5\times 10^{-6}$.
The multifractal analysis of the conductance series is shown in Figs.~\ref{Hq_e_falfa_pontos}(a) and \ref{Hq_e_falfa_pontos}(b), where the mean values of $H(p)$ and $f(\alpha)$ were obtained from 10 different conductance series, changing the billiard boundaries.
Figures \ref{Hq_e_falfa_pontos}(a) and \ref{Hq_e_falfa_pontos}(b) have the same multifractal behavior observed in disordered nanowires (Fig.~\ref{Hq_e_falfa_fios}).
This means that both dependence of $H(p)$ on $p$ and the value of $\Delta\alpha$ decrease when $N$ increases, which indicates a weakening of the multifractal behavior, as expected.

We have also performed electronic transport calculations of chaotic billiards modeled as confined two-dimensional electron gases with a square lattice (2DCB). The conductance series were obtained tuning the Fermi energy in order to set the number of propagating modes in the leads to $N=1,2,5,10$, and using magnetic flux increments $\Delta\phi = 1\times 10^{-5}$.
The mean values of $H(p)$ and $f(\alpha)$ obtained from 10 different conductance series are shown in Figs.~\ref{Hq_e_falfa_pontos}(c) and \ref{Hq_e_falfa_pontos}(d), respectively. 
It is quite clear that 2DCB show the same multifractal behavior observed both in disordered nanowires (Fig.~\ref{Hq_e_falfa_fios}) and GCB, as can be seen in Figs.~\ref{Hq_e_falfa_pontos}(a) and \ref{Hq_e_falfa_pontos}(b). In this case, one can clearly see a crossover from multifractal to monofractal behavior as $N$ increases, since $H(p)$ becomes approximately independent of $p$ (which can be seen by a flattening of curve $H(p)$ when $N$ changes from $1$ to $10$) and $\Delta\alpha \rightarrow 0$ when $N$ increases.

\section{Distribution of Conductance Increments}\label{results_distributions}

In order to gain a deeper understanding of MCF for all mesoscopic devices considered in this paper, we studied the distribution of conductance increments, given by Eq.~(\ref{GB}), which can be seen as a measure of the electronic transport response due to a small magnetic field perturbation. Besides, conductance increments can carry information about  certain quantum correlations (induced by the magnetic field) inside the scattering region.

\begin{figure}
    \centering
    \includegraphics[width=0.9\linewidth]{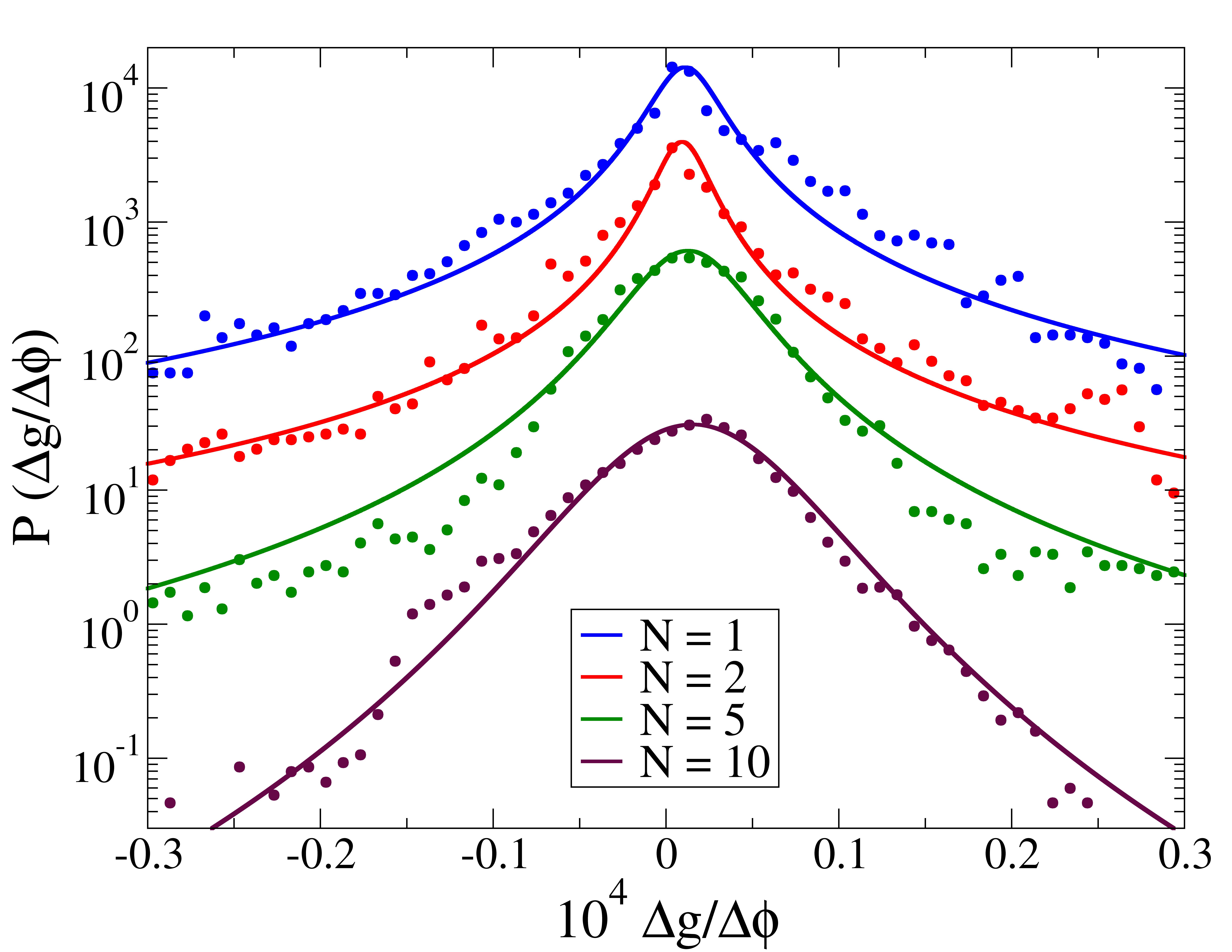}
    \caption{NCI histograms of MCF data of graphene chaotic billiards divided by the flux step $\Delta\phi$ with $N=1$ (blue squares), $N=2$ (red squares), $N=5$ (green squares), $N=10$ (maroon squares) propagating channels, which were obtained from the same series we have used to perform the multifractal analysis in Sec. \ref{sec_results}. The solid lines show the best fits with a $q$-Gaussian function. Notice that the peak of the distribution becomes narrower as $N$ decreases. The parameter values of the $q$-Gaussian functions which best fit the points are $q=2.09 \pm0.05, 2.09\pm0.06, 1.73\pm0.05, 1.34\pm0.07$, $\beta=2375 \pm335, 4022 \pm643, 952 \pm79, 364 \pm31$ and $x_0=0.011 \pm0.001, 0.009 \pm0.001, 0.013 \pm0.001, 0.016 \pm0.001$, for $N=1,2,5,10$, respectively. The curves have been  arbitrarily shifted in the vertical direction for clarity.}
    \label{fits_q_gaussian_ponto_grafeno}
\end{figure}

For convenience, we introduce the normalized conductance increments (NCI), $x={\Delta g}/{\Delta \phi}$, where ${\Delta g}={g(\phi+\Delta\phi)-g(\phi)}$, for a fixed Fermi energy.  
This normalization only widens the tails of the distribution without loss of information. 
Figure \ref{fits_q_gaussian_ponto_grafeno} shows NCI histograms from the conductance series of GCB with $N=1,2,5,10$ propagating modes in the leads (filled circles). To build the NCI histograms for the GCB, we have combined all 10 realizations of this system; see Sec.~\ref{sec:GCB}.


Interestingly, we observe that one can fit well all NCI histograms with the $q$-Gaussian probability density function:
\begin{equation}\label{q-gaussian_equation}
    P(x) = \frac{\sqrt{\beta}}{C_q}[1+(q-1)\beta (x-x_0)^{2}]^{\frac{1}{1-q}},
\end{equation}
with
\begin{equation*}
   C_q = \frac{\sqrt{\pi}\Gamma\Big(\frac{3-q}{2(q-1)}\Big)}{\sqrt{q-1}\Gamma\Big(\frac{1}{q-1}\Big)},
\end{equation*}
where $1<q<3$, $\beta$ is a measure of the width of the distribution, and $x_0$ is its mean.
We remark that Eq.~(\ref{q-gaussian_equation}) can be formally derived from a maximization of the Tsallis entropy \cite{q-statistics}. Note that, when $q\rightarrow 1 $, $P(x)$ converges to the Gaussian distribution. Therefore, values of $q$ different from 1 can be seen as a measure of non-Gaussianity. In Fig.~\ref{fits_q_gaussian_ponto_grafeno} we show the NCI histograms (filled circles), together with the best fit by $q$-Gaussian functions (solid lines), with $q$ values given by $q=2.09 \pm0.05, 2.09\pm0.06, 1.73\pm0.05, 1.34\pm0.07$, for $N=1,2,5,10$, respectively. (In Fig.~\ref{fits_q_gaussian_ponto_grafeno} the curves have been  arbitrarily shifted in the vertical direction for clarity.)

In Fig.~\ref{fits_q_gaussian_ponto_grafeno}, we observe two important features of the extreme quantum regime ($N=1$), namely the distributions have {\it heavy tails} and {\it sharp peaks} around zero, which means that they are highly non-Gaussian. Furthermore, as $N$ increases, both the heavy tails and the sharp peaks tend to become less pronounced, thus indicating that the NCI distributions perform a crossover from a non-Gaussian behavior in the extreme quantum regime ($N=1$) to a Gaussian behavior in the semiclassical regime ($N\gg1$).
This behavior can be related to a gradual loss of correlations induced by the magnetic field in the stochastic process associated with the conductance series as $N$ becomes large \cite{frahm-pichard1},
which can be quantified by a decrease  of $q$ as $N$ increases. In other words, we expect $q\to1$ for $N\gg1$, as seen in the trend observed in  Fig.~\ref{fits_q_gaussian_ponto_grafeno}.
Finally, we observed similar behaviors for all NCI distributions in the other mesoscopic devices considered in this paper: AGNR, ZGNR, 2DEG, and 2DCB, as will be discussed in Sec.~\ref{discussion}.

\section{Discussion of the results}\label{discussion}


Experimental observations of multifractality in the conductance fluctuations of single-layer graphene were reported in Ref.~\cite{Amin2018}. 
There the authors showed that conductance fluctuations originated from the variation of an applied magnetic field to a high-mobility single-layer graphene exhibit multifractal scaling under two conditions, namely very low temperature and proximity to the charge-neutrality (Dirac) point, i.e., the extreme quantum regime. 
They suggested that the experimental measurements and analysis presented evidence of an incipient Anderson localization near the Dirac point as the most plausible cause for this multifractality. Multifractality in disordered quantum systems is indeed observed, for example, in the scaling of the eigenfunctions in the vicinity of an Anderson transition \cite{anderson-transitions}; and the key idea presented in \cite{Amin2018} was that the wave function multifractal behavior is simply transferred to multifractality in conductance fluctuations.

In the present study, we have performed a multifractal analysis of the magnetic flux-induced conductance fluctuations in three types of disordered nanowires (Fig.~\ref{Hq_e_falfa_fios}) and two different ballistic chaotic billiards (Fig.~\ref{Hq_e_falfa_pontos}), and observed that in the quantum regime they all exhibit multifractal scaling. The first consequence of our study is that the multifractal behavior is not a specific feature of single-layer graphene, but a generic feature of the quantum regime of fluctuations.  Second, our results give evidence that MCF are not necessarily related to an incipient Anderson localization, since there is no localization effect in ballistic chaotic billiards. 
We suggest that MCF are rather a consequence of the ubiquitous quantum-mechanical interference which characterizes mesoscopic phenomena. More specifically, we argue that MCF are caused by correlations induced by the external magnetic field in the stochastic process associated with the conductance series. 
Interestingly, there is a very natural interpretation for correlations (induced by the magnetic field) that are strong in the quantum regime and are gradually lost when the system enters into the semiclassical regime of conduction. In the language of the trajectory-based semiclassical approach, they are described as correlations between semiclassical trajectories near close encounters \cite{PhysRevLett.93.014103, Novaes_2013, PhysRevLett.89.206801,PhysRevE.85.045201,Kuipers_2013}. 

Moreover, from the results given in Sec.~\ref{sec_results}, we extract three arguments to sustain the notion that multifractality is a generic feature of conductance fluctuations of mesoscopic systems in the quantum regime. First, the conductance fluctuations of both disordered and ballistic systems are multifractal in the quantum regime, as can be observed if one compares the results in Figs.~\ref{Hq_e_falfa_fios} and \ref{Hq_e_falfa_pontos}. Second, the fluctuations are multifractal for both square and hexagonal lattice structures, which leads us to believe that, for systems made of material other than graphene, results similar to those presented in \cite{Amin2018} may be observed. Last, in the case of graphene nanoribbons, the multifractal nature of the conductance fluctuations of the system is independent of the geometry of the boundaries, since the results presented in the top [(a), (b)] and middle [(c), (d)] panels of Fig.~\ref{Hq_e_falfa_fios} for graphene nanoribbons with armchair and zigzag boundaries, respectively, are equivalent.

\begin{figure*}
    \centering
    \includegraphics[width=0.95\linewidth]{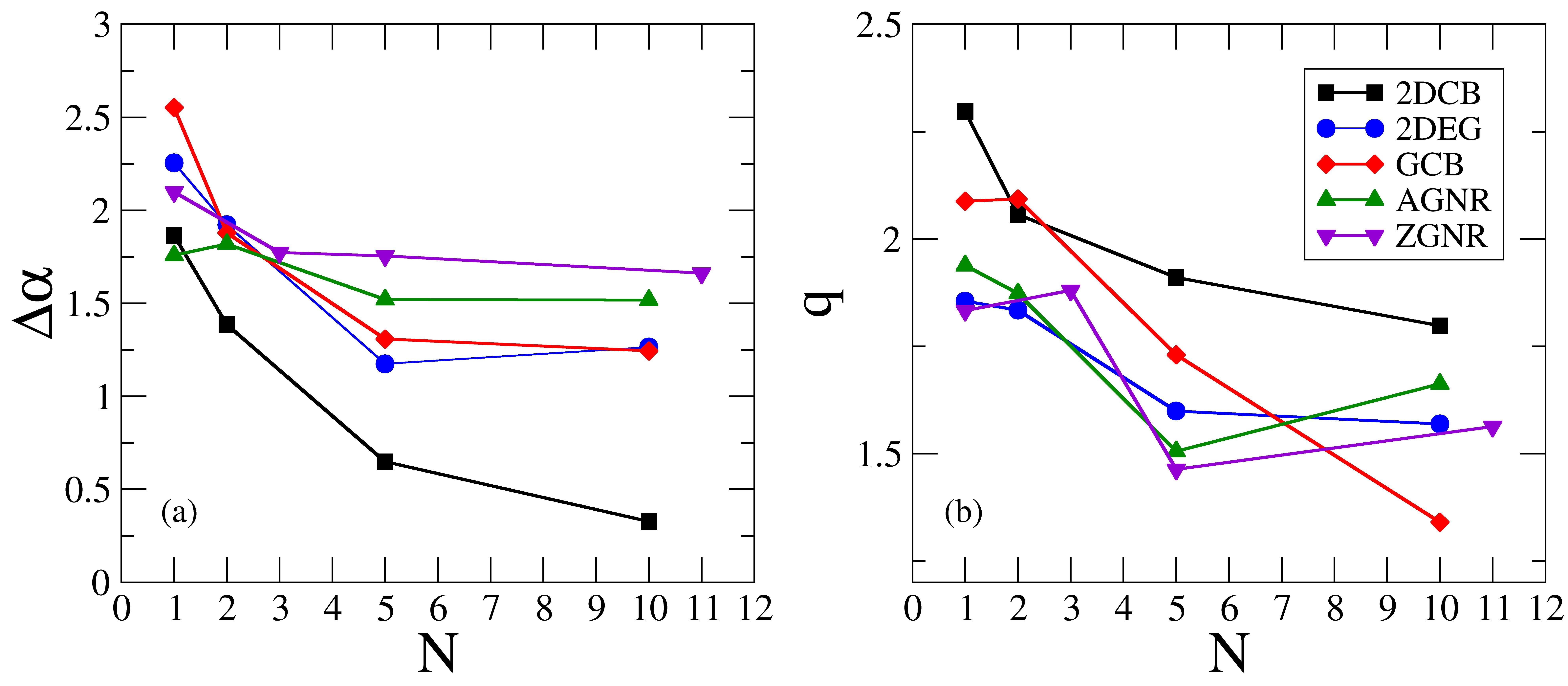}
    \caption{Dependence of (a) the width of the multifractal singularity spectrum $f(\alpha)$ with $N$ and of (b) the value of $q$ that best fits the data of the distribution of conductance increments divided by $\Delta\phi$. The black squares, blue circles, red diamonds, green up triangles, and violet down triangles represent the results for two-dimensional chaotic billiards with a square lattice (2DCB), two-dimensional electron gases with a square lattice (2DEG), graphene chaotic billiards (GCB), armchair graphene nanoribbons (AGNR), and zigzag graphene nanoribbons (ZGNR), respectively.}
    \label{Delta_alfa_e_q_vs_N}
\end{figure*}

The central result of this paper is shown in Fig.~\ref{Delta_alfa_e_q_vs_N}. On the left panel, we plot $\Delta \alpha$ as a function of $N$. Although the conductance fluctuations of both ballistic chaotic billiards and disordered nanowires are multifractal in the extreme quantum regime ($N=1$), there is a weakening of the multifractal behavior as one increases the number $N$ of propagating modes in the leads, which is 
evidenced by a corresponding decrease in the width $\Delta \alpha$ with the increase of $N$ for all the systems considered.
As it is well known, the presence of correlations in time series is one of the possible origins of multifractality \cite{mfdfa}, and this source can be tracked by performing the multifractal analysis of the shuffled time series. Indeed, we have performed the multifractal analysis of the shuffled magnetic field-induced conductance fluctuations in one of the 10 realizations of 2DCB (square lattice) and observed that the generalized Hurst exponent $H(p)$ is approximately independent of $p$, with value $H(p)\simeq0.5$, which means that the shuffled series is monofractal, and this happens because all correlations previously present are erased with the shuffling procedure.

The presence or absence of correlations in our conductance fictitious-time series can also be studied from the point of view of the distribution of conductance increments, as the ones shown in Fig.~\ref{fits_q_gaussian_ponto_grafeno}. This distribution is expected to be Gaussian when the series are completely uncorrelated. In Sec.~\ref{results_distributions}, we remarked that all NCI distributions can be fitted by a $q$-Gaussian distribution, given by Eq.~(\ref{q-gaussian_equation}). The values of $q$ are shown in the right panel of Fig.~\ref{Delta_alfa_e_q_vs_N} as functions of $N$. We can  clearly see both $\Delta \alpha$ and $q$ decrease with $N$, which means that the loss of multifractal scaling can be directly associated with the loss of  correlation of the conductance time series.

One interesting final remark is that the $q$-Gaussian function is a special case of the class of universal functions which can be used to describe hierarchical systems; specifically it corresponds to the case of systems with dynamical variables with probability density functions with power-law tails and whose stochastic dynamics has a single hierarchical level \cite{salazar-vasconcelos,pre-H}. With this in mind, we conclude that MCF can be thought of as related to the hierarchical nature of the stochastic process of the conductance in terms of the magnetic field, which opens a different point of view to explore multifractality in mesoscopic phenomena.

\section{Conclusions}

We have observed that the magnetic field-induced conductance fluctuations of mesoscopic conductors, which were numerically obtained via a tight-binding approach, are multifractal. Our  analysis led us to conclude that this multifractality is not originated by an incipient Anderson transition, as pointed out in \cite{Amin2018}, where an experimental observation of MCF was reported, but rather by statistical correlations induced by the externally applied magnetic field in the stochastic process of the corresponding conductance series. Additionally, we have shown that the distribution of conductance increments, or magnetoconductance, is well fitted by the $q$-Gaussian function, which is a very important heavy-tailed distribution with many applications. It remains an open possibility to study the MCF generated by changing other parameters, such as the Fermi energy of the system.

\begin{acknowledgments}
 This work was supported in part by CNPq and CAPES (Brazilian agencies).  
\end{acknowledgments}

\bibliography{apssamp}

\providecommand{\noopsort}[1]{}\providecommand{\singleletter}[1]{#1}%
\begin{thebibliography}{33}%
\makeatletter
\providecommand \@ifxundefined [1]{%
 \@ifx{#1\undefined}
}%
\providecommand \@ifnum [1]{%
 \ifnum #1\expandafter \@firstoftwo
 \else \expandafter \@secondoftwo
 \fi
}%
\providecommand \@ifx [1]{%
 \ifx #1\expandafter \@firstoftwo
 \else \expandafter \@secondoftwo
 \fi
}%
\providecommand \natexlab [1]{#1}%
\providecommand \enquote  [1]{``#1''}%
\providecommand \bibnamefont  [1]{#1}%
\providecommand \bibfnamefont [1]{#1}%
\providecommand \citenamefont [1]{#1}%
\providecommand \href@noop [0]{\@secondoftwo}%
\providecommand \href [0]{\begingroup \@sanitize@url \@href}%
\providecommand \@href[1]{\@@startlink{#1}\@@href}%
\providecommand \@@href[1]{\endgroup#1\@@endlink}%
\providecommand \@sanitize@url [0]{\catcode `\\12\catcode `\$12\catcode
  `\&12\catcode `\#12\catcode `\^12\catcode `\_12\catcode `\%12\relax}%
\providecommand \@@startlink[1]{}%
\providecommand \@@endlink[0]{}%
\providecommand \url  [0]{\begingroup\@sanitize@url \@url }%
\providecommand \@url [1]{\endgroup\@href {#1}{\urlprefix }}%
\providecommand \urlprefix  [0]{URL }%
\providecommand \Eprint [0]{\href }%
\providecommand \doibase [0]{https://doi.org/}%
\providecommand \selectlanguage [0]{\@gobble}%
\providecommand \bibinfo  [0]{\@secondoftwo}%
\providecommand \bibfield  [0]{\@secondoftwo}%
\providecommand \translation [1]{[#1]}%
\providecommand \BibitemOpen [0]{}%
\providecommand \bibitemStop [0]{}%
\providecommand \bibitemNoStop [0]{.\EOS\space}%
\providecommand \EOS [0]{\spacefactor3000\relax}%
\providecommand \BibitemShut  [1]{\csname bibitem#1\endcsname}%
\let\auto@bib@innerbib\@empty
\bibitem [{\citenamefont {Beenakker}(1997)}]{RevModPhys.69.731}%
  \BibitemOpen
  \bibfield  {author} {\bibinfo {author} {\bibfnamefont {C.~W.~J.}\
  \bibnamefont {Beenakker}},\ }\bibfield  {title} {\bibinfo {title}
  {Random-matrix theory of quantum transport},\ }\href
  {https://doi.org/10.1103/RevModPhys.69.731} {\bibfield  {journal} {\bibinfo
  {journal} {Rev. Mod. Phys.}\ }\textbf {\bibinfo {volume} {69}},\ \bibinfo
  {pages} {731} (\bibinfo {year} {1997})}\BibitemShut {NoStop}%
\bibitem [{\citenamefont {Rotter}\ and\ \citenamefont
  {Gigan}(2017)}]{RevModPhys.89.015005}%
  \BibitemOpen
  \bibfield  {author} {\bibinfo {author} {\bibfnamefont {S.}~\bibnamefont
  {Rotter}}\ and\ \bibinfo {author} {\bibfnamefont {S.}~\bibnamefont {Gigan}},\
  }\bibfield  {title} {\bibinfo {title} {Light fields in complex media:
  Mesoscopic scattering meets wave control},\ }\href
  {https://doi.org/10.1103/RevModPhys.89.015005} {\bibfield  {journal}
  {\bibinfo  {journal} {Rev. Mod. Phys.}\ }\textbf {\bibinfo {volume} {89}},\
  \bibinfo {pages} {015005} (\bibinfo {year} {2017})}\BibitemShut {NoStop}%
\bibitem [{\citenamefont {Hegger}\ \emph {et~al.}(1996)\citenamefont {Hegger},
  \citenamefont {Huckestein}, \citenamefont {Hecker}, \citenamefont {Janssen},
  \citenamefont {Freimuth}, \citenamefont {Reckziegel},\ and\ \citenamefont
  {Tuzinski}}]{gold}%
  \BibitemOpen
  \bibfield  {author} {\bibinfo {author} {\bibfnamefont {H.}~\bibnamefont
  {Hegger}}, \bibinfo {author} {\bibfnamefont {B.}~\bibnamefont {Huckestein}},
  \bibinfo {author} {\bibfnamefont {K.}~\bibnamefont {Hecker}}, \bibinfo
  {author} {\bibfnamefont {M.}~\bibnamefont {Janssen}}, \bibinfo {author}
  {\bibfnamefont {A.}~\bibnamefont {Freimuth}}, \bibinfo {author}
  {\bibfnamefont {G.}~\bibnamefont {Reckziegel}},\ and\ \bibinfo {author}
  {\bibfnamefont {R.}~\bibnamefont {Tuzinski}},\ }\bibfield  {title} {\bibinfo
  {title} {Fractal conductance fluctuations in gold nanowires},\ }\href
  {https://doi.org/10.1103/PhysRevLett.77.3885} {\bibfield  {journal} {\bibinfo
   {journal} {Phys. Rev. Lett.}\ }\textbf {\bibinfo {volume} {77}},\ \bibinfo
  {pages} {3885} (\bibinfo {year} {1996})}\BibitemShut {NoStop}%
\bibitem [{\citenamefont {Ver\ifmmode~\mbox{\c{c}}\else \c{c}\fi{}osa}\ \emph
  {et~al.}(2018)\citenamefont {Ver\ifmmode~\mbox{\c{c}}\else \c{c}\fi{}osa},
  \citenamefont {Doh}, \citenamefont {Ramos},\ and\ \citenamefont
  {Barbosa}}]{PhysRevB.98.155407}%
  \BibitemOpen
  \bibfield  {author} {\bibinfo {author} {\bibfnamefont {T.}~\bibnamefont
  {Ver\ifmmode~\mbox{\c{c}}\else \c{c}\fi{}osa}}, \bibinfo {author}
  {\bibfnamefont {Y.-J.}\ \bibnamefont {Doh}}, \bibinfo {author} {\bibfnamefont
  {J.~G. G.~S.}\ \bibnamefont {Ramos}},\ and\ \bibinfo {author} {\bibfnamefont
  {A.~L.~R.}\ \bibnamefont {Barbosa}},\ }\bibfield  {title} {\bibinfo {title}
  {Conductance peak density in nanowires},\ }\href
  {https://doi.org/10.1103/PhysRevB.98.155407} {\bibfield  {journal} {\bibinfo
  {journal} {Phys. Rev. B}\ }\textbf {\bibinfo {volume} {98}},\ \bibinfo
  {pages} {155407} (\bibinfo {year} {2018})}\BibitemShut {NoStop}%
\bibitem [{\citenamefont {Santana}\ \emph {et~al.}(2020)\citenamefont
  {Santana}, \citenamefont {da~Silva}, \citenamefont {Vasconcelos},
  \citenamefont {Ramos},\ and\ \citenamefont {Barbosa}}]{PhysRevB.102.041107}%
  \BibitemOpen
  \bibfield  {author} {\bibinfo {author} {\bibfnamefont {F.~A.~F.}\
  \bibnamefont {Santana}}, \bibinfo {author} {\bibfnamefont {J.~M.}\
  \bibnamefont {da~Silva}}, \bibinfo {author} {\bibfnamefont {T.~C.}\
  \bibnamefont {Vasconcelos}}, \bibinfo {author} {\bibfnamefont {J.~G. G.~S.}\
  \bibnamefont {Ramos}},\ and\ \bibinfo {author} {\bibfnamefont {A.~L.~R.}\
  \bibnamefont {Barbosa}},\ }\bibfield  {title} {\bibinfo {title} {Spin hall
  angle fluctuations in a device with disorder},\ }\href
  {https://doi.org/10.1103/PhysRevB.102.041107} {\bibfield  {journal} {\bibinfo
   {journal} {Phys. Rev. B}\ }\textbf {\bibinfo {volume} {102}},\ \bibinfo
  {pages} {041107} (\bibinfo {year} {2020})}\BibitemShut {NoStop}%
\bibitem [{\citenamefont {S\'a}\ \emph {et~al.}(2020)\citenamefont {S\'a},
  \citenamefont {Barbosa},\ and\ \citenamefont {Ramos}}]{PhysRevB.102.115105}%
  \BibitemOpen
  \bibfield  {author} {\bibinfo {author} {\bibfnamefont {L.~G. C.~S.}\
  \bibnamefont {S\'a}}, \bibinfo {author} {\bibfnamefont {A.~L.~R.}\
  \bibnamefont {Barbosa}},\ and\ \bibinfo {author} {\bibfnamefont {J.~G.
  G.~S.}\ \bibnamefont {Ramos}},\ }\bibfield  {title} {\bibinfo {title}
  {Conductance peak density in disordered graphene topological insulators},\
  }\href {https://doi.org/10.1103/PhysRevB.102.115105} {\bibfield  {journal}
  {\bibinfo  {journal} {Phys. Rev. B}\ }\textbf {\bibinfo {volume} {102}},\
  \bibinfo {pages} {115105} (\bibinfo {year} {2020})}\BibitemShut {NoStop}%
\bibitem [{\citenamefont {{\hspace{0.167em}}V~Brazhkin}\ and\ \citenamefont
  {Suslov}(2020)}]{Brazhkin_2020}%
  \BibitemOpen
  \bibfield  {author} {\bibinfo {author} {\bibfnamefont {V.}~\bibnamefont
  {{\hspace{0.167em}}V~Brazhkin}}\ and\ \bibinfo {author} {\bibfnamefont
  {I.~M.}\ \bibnamefont {Suslov}},\ }\bibfield  {title} {\bibinfo {title}
  {Mechanism of universal conductance fluctuations},\ }\href
  {https://doi.org/10.1088/1361-648x/ab8ec5} {\bibfield  {journal} {\bibinfo
  {journal} {Journal of Physics: Condensed Matter}\ }\textbf {\bibinfo {volume}
  {32}},\ \bibinfo {pages} {35LT02} (\bibinfo {year} {2020})}\BibitemShut
  {NoStop}%
\bibitem [{\citenamefont {Sachrajda}\ \emph {et~al.}(1998)\citenamefont
  {Sachrajda}, \citenamefont {Ketzmerick}, \citenamefont {Gould}, \citenamefont
  {Feng}, \citenamefont {Kelly}, \citenamefont {Delage},\ and\ \citenamefont
  {Wasilewski}}]{sachrajda}%
  \BibitemOpen
  \bibfield  {author} {\bibinfo {author} {\bibfnamefont {A.~S.}\ \bibnamefont
  {Sachrajda}}, \bibinfo {author} {\bibfnamefont {R.}~\bibnamefont
  {Ketzmerick}}, \bibinfo {author} {\bibfnamefont {C.}~\bibnamefont {Gould}},
  \bibinfo {author} {\bibfnamefont {Y.}~\bibnamefont {Feng}}, \bibinfo {author}
  {\bibfnamefont {P.~J.}\ \bibnamefont {Kelly}}, \bibinfo {author}
  {\bibfnamefont {A.}~\bibnamefont {Delage}},\ and\ \bibinfo {author}
  {\bibfnamefont {Z.}~\bibnamefont {Wasilewski}},\ }\bibfield  {title}
  {\bibinfo {title} {Fractal conductance fluctuations in a soft-wall stadium
  and a sinai billiard},\ }\href {https://doi.org/10.1103/PhysRevLett.80.1948}
  {\bibfield  {journal} {\bibinfo  {journal} {Phys. Rev. Lett.}\ }\textbf
  {\bibinfo {volume} {80}},\ \bibinfo {pages} {1948} (\bibinfo {year}
  {1998})}\BibitemShut {NoStop}%
\bibitem [{\citenamefont {Taylor}\ \emph
  {et~al.}(1997{\natexlab{a}})\citenamefont {Taylor}, \citenamefont {Newbury},
  \citenamefont {Sachrajda}, \citenamefont {Feng}, \citenamefont {Coleridge},
  \citenamefont {Dettmann}, \citenamefont {Zhu}, \citenamefont {Guo},
  \citenamefont {Delage}, \citenamefont {Kelly},\ and\ \citenamefont
  {Wasilewski}}]{taylor1}%
  \BibitemOpen
  \bibfield  {author} {\bibinfo {author} {\bibfnamefont {R.~P.}\ \bibnamefont
  {Taylor}}, \bibinfo {author} {\bibfnamefont {R.}~\bibnamefont {Newbury}},
  \bibinfo {author} {\bibfnamefont {A.~S.}\ \bibnamefont {Sachrajda}}, \bibinfo
  {author} {\bibfnamefont {Y.}~\bibnamefont {Feng}}, \bibinfo {author}
  {\bibfnamefont {P.~T.}\ \bibnamefont {Coleridge}}, \bibinfo {author}
  {\bibfnamefont {C.}~\bibnamefont {Dettmann}}, \bibinfo {author}
  {\bibfnamefont {N.}~\bibnamefont {Zhu}}, \bibinfo {author} {\bibfnamefont
  {H.}~\bibnamefont {Guo}}, \bibinfo {author} {\bibfnamefont {A.}~\bibnamefont
  {Delage}}, \bibinfo {author} {\bibfnamefont {P.~J.}\ \bibnamefont {Kelly}},\
  and\ \bibinfo {author} {\bibfnamefont {Z.}~\bibnamefont {Wasilewski}},\
  }\bibfield  {title} {\bibinfo {title} {Self-similar magnetoresistance of a
  semiconductor sinai billiard},\ }\href
  {https://doi.org/10.1103/PhysRevLett.78.1952} {\bibfield  {journal} {\bibinfo
   {journal} {Phys. Rev. Lett.}\ }\textbf {\bibinfo {volume} {78}},\ \bibinfo
  {pages} {1952} (\bibinfo {year} {1997}{\natexlab{a}})}\BibitemShut {NoStop}%
\bibitem [{\citenamefont {Taylor}\ \emph
  {et~al.}(1997{\natexlab{b}})\citenamefont {Taylor}, \citenamefont {Micolich},
  \citenamefont {Newbury},\ and\ \citenamefont {Fromhold}}]{taylor2}%
  \BibitemOpen
  \bibfield  {author} {\bibinfo {author} {\bibfnamefont {R.~P.}\ \bibnamefont
  {Taylor}}, \bibinfo {author} {\bibfnamefont {A.~P.}\ \bibnamefont
  {Micolich}}, \bibinfo {author} {\bibfnamefont {R.}~\bibnamefont {Newbury}},\
  and\ \bibinfo {author} {\bibfnamefont {T.~M.}\ \bibnamefont {Fromhold}},\
  }\bibfield  {title} {\bibinfo {title} {Correlation analysis of
  self-similarity in semiconductor billiards},\ }\href
  {https://doi.org/10.1103/PhysRevB.56.R12733} {\bibfield  {journal} {\bibinfo
  {journal} {Phys. Rev. B}\ }\textbf {\bibinfo {volume} {56}},\ \bibinfo
  {pages} {R12733} (\bibinfo {year} {1997}{\natexlab{b}})}\BibitemShut
  {NoStop}%
\bibitem [{\citenamefont {Taylor}\ \emph {et~al.}(1998)\citenamefont {Taylor},
  \citenamefont {Micolich}, \citenamefont {Newbury}, \citenamefont {Bird},
  \citenamefont {Fromhold}, \citenamefont {Cooper}, \citenamefont {Aoyagi},\
  and\ \citenamefont {Sugano}}]{taylor3}%
  \BibitemOpen
  \bibfield  {author} {\bibinfo {author} {\bibfnamefont {R.~P.}\ \bibnamefont
  {Taylor}}, \bibinfo {author} {\bibfnamefont {A.~P.}\ \bibnamefont
  {Micolich}}, \bibinfo {author} {\bibfnamefont {R.}~\bibnamefont {Newbury}},
  \bibinfo {author} {\bibfnamefont {J.~P.}\ \bibnamefont {Bird}}, \bibinfo
  {author} {\bibfnamefont {T.~M.}\ \bibnamefont {Fromhold}}, \bibinfo {author}
  {\bibfnamefont {J.}~\bibnamefont {Cooper}}, \bibinfo {author} {\bibfnamefont
  {Y.}~\bibnamefont {Aoyagi}},\ and\ \bibinfo {author} {\bibfnamefont
  {T.}~\bibnamefont {Sugano}},\ }\bibfield  {title} {\bibinfo {title} {Exact
  and statistical self-similarity in magnetoconductance fluctuations: A unified
  picture},\ }\href {https://doi.org/10.1103/PhysRevB.58.11107} {\bibfield
  {journal} {\bibinfo  {journal} {Phys. Rev. B}\ }\textbf {\bibinfo {volume}
  {58}},\ \bibinfo {pages} {11107} (\bibinfo {year} {1998})}\BibitemShut
  {NoStop}%
\bibitem [{\citenamefont {Wurm}\ \emph {et~al.}(2009)\citenamefont {Wurm},
  \citenamefont {Rycerz}, \citenamefont {Adagideli}, \citenamefont {Wimmer},
  \citenamefont {Richter},\ and\ \citenamefont {Baranger}}]{halfstadium}%
  \BibitemOpen
  \bibfield  {author} {\bibinfo {author} {\bibfnamefont {J.}~\bibnamefont
  {Wurm}}, \bibinfo {author} {\bibfnamefont {A.}~\bibnamefont {Rycerz}},
  \bibinfo {author} {\bibfnamefont {i.~d. I. m.~c.}\ \bibnamefont {Adagideli}},
  \bibinfo {author} {\bibfnamefont {M.}~\bibnamefont {Wimmer}}, \bibinfo
  {author} {\bibfnamefont {K.}~\bibnamefont {Richter}},\ and\ \bibinfo {author}
  {\bibfnamefont {H.~U.}\ \bibnamefont {Baranger}},\ }\bibfield  {title}
  {\bibinfo {title} {Symmetry classes in graphene quantum dots: Universal
  spectral statistics, weak localization, and conductance fluctuations},\
  }\href {https://doi.org/10.1103/PhysRevLett.102.056806} {\bibfield  {journal}
  {\bibinfo  {journal} {Phys. Rev. Lett.}\ }\textbf {\bibinfo {volume} {102}},\
  \bibinfo {pages} {056806} (\bibinfo {year} {2009})}\BibitemShut {NoStop}%
\bibitem [{\citenamefont {Barbosa}\ \emph {et~al.}(2021)\citenamefont
  {Barbosa}, \citenamefont {Ramos},\ and\ \citenamefont
  {Ferreira}}]{halfstadium2}%
  \BibitemOpen
  \bibfield  {author} {\bibinfo {author} {\bibfnamefont {A.~L.~R.}\
  \bibnamefont {Barbosa}}, \bibinfo {author} {\bibfnamefont {J.~G. G.~S.}\
  \bibnamefont {Ramos}},\ and\ \bibinfo {author} {\bibfnamefont
  {A.}~\bibnamefont {Ferreira}},\ }\bibfield  {title} {\bibinfo {title} {Effect
  of proximity-induced spin-orbit coupling in graphene mesoscopic billiards},\
  }\href {https://doi.org/10.1103/PhysRevB.103.L081111} {\bibfield  {journal}
  {\bibinfo  {journal} {Phys. Rev. B}\ }\textbf {\bibinfo {volume} {103}},\
  \bibinfo {pages} {L081111} (\bibinfo {year} {2021})}\BibitemShut {NoStop}%
\bibitem [{\citenamefont {Huang}\ \emph {et~al.}(2018)\citenamefont {Huang},
  \citenamefont {Xu}, \citenamefont {Grebogi},\ and\ \citenamefont
  {Lai}}]{HUANG20181}%
  \BibitemOpen
  \bibfield  {author} {\bibinfo {author} {\bibfnamefont {L.}~\bibnamefont
  {Huang}}, \bibinfo {author} {\bibfnamefont {H.-Y.}\ \bibnamefont {Xu}},
  \bibinfo {author} {\bibfnamefont {C.}~\bibnamefont {Grebogi}},\ and\ \bibinfo
  {author} {\bibfnamefont {Y.-C.}\ \bibnamefont {Lai}},\ }\bibfield  {title}
  {\bibinfo {title} {Relativistic quantum chaos},\ }\href
  {https://doi.org/https://doi.org/10.1016/j.physrep.2018.06.006} {\bibfield
  {journal} {\bibinfo  {journal} {Physics Reports}\ }\textbf {\bibinfo {volume}
  {753}},\ \bibinfo {pages} {1} (\bibinfo {year} {2018})},\ \bibinfo {note}
  {relativistic Quantum Chaos}\BibitemShut {NoStop}%
\bibitem [{\citenamefont {Washburn}\ and\ \citenamefont
  {Webb}(1986)}]{washburn}%
  \BibitemOpen
  \bibfield  {author} {\bibinfo {author} {\bibfnamefont {S.}~\bibnamefont
  {Washburn}}\ and\ \bibinfo {author} {\bibfnamefont {R.~A.}\ \bibnamefont
  {Webb}},\ }\bibfield  {title} {\bibinfo {title} {Aharonov-bohm effect in
  normal metal quantum coherence and transport},\ }\href
  {https://doi.org/10.1080/00018738600101921} {\bibfield  {journal} {\bibinfo
  {journal} {Advances in Physics}\ }\textbf {\bibinfo {volume} {35}},\ \bibinfo
  {pages} {375} (\bibinfo {year} {1986})}\BibitemShut {NoStop}%
\bibitem [{\citenamefont {Lee}\ and\ \citenamefont
  {Stone}(1985)}]{UCF_Lee_Stone}%
  \BibitemOpen
  \bibfield  {author} {\bibinfo {author} {\bibfnamefont {P.~A.}\ \bibnamefont
  {Lee}}\ and\ \bibinfo {author} {\bibfnamefont {A.~D.}\ \bibnamefont
  {Stone}},\ }\bibfield  {title} {\bibinfo {title} {Universal conductance
  fluctuations in metals},\ }\href
  {https://doi.org/10.1103/PhysRevLett.55.1622} {\bibfield  {journal} {\bibinfo
   {journal} {Phys. Rev. Lett.}\ }\textbf {\bibinfo {volume} {55}},\ \bibinfo
  {pages} {1622} (\bibinfo {year} {1985})}\BibitemShut {NoStop}%
\bibitem [{\citenamefont {Lee}\ \emph {et~al.}(1987)\citenamefont {Lee},
  \citenamefont {Stone},\ and\ \citenamefont
  {Fukuyama}}]{UCF_Lee_Stone_Fukuyama}%
  \BibitemOpen
  \bibfield  {author} {\bibinfo {author} {\bibfnamefont {P.~A.}\ \bibnamefont
  {Lee}}, \bibinfo {author} {\bibfnamefont {A.~D.}\ \bibnamefont {Stone}},\
  and\ \bibinfo {author} {\bibfnamefont {H.}~\bibnamefont {Fukuyama}},\
  }\bibfield  {title} {\bibinfo {title} {Universal conductance fluctuations in
  metals: Effects of finite temperature, interactions, and magnetic field},\
  }\href {https://doi.org/10.1103/PhysRevB.35.1039} {\bibfield  {journal}
  {\bibinfo  {journal} {Phys. Rev. B}\ }\textbf {\bibinfo {volume} {35}},\
  \bibinfo {pages} {1039} (\bibinfo {year} {1987})}\BibitemShut {NoStop}%
\bibitem [{\citenamefont {Jalabert}\ \emph {et~al.}(1990)\citenamefont
  {Jalabert}, \citenamefont {Baranger},\ and\ \citenamefont
  {Stone}}]{jalabert}%
  \BibitemOpen
  \bibfield  {author} {\bibinfo {author} {\bibfnamefont {R.~A.}\ \bibnamefont
  {Jalabert}}, \bibinfo {author} {\bibfnamefont {H.~U.}\ \bibnamefont
  {Baranger}},\ and\ \bibinfo {author} {\bibfnamefont {A.~D.}\ \bibnamefont
  {Stone}},\ }\bibfield  {title} {\bibinfo {title} {Conductance fluctuations in
  the ballistic regime: A probe of quantum chaos?},\ }\href
  {https://doi.org/10.1103/PhysRevLett.65.2442} {\bibfield  {journal} {\bibinfo
   {journal} {Phys. Rev. Lett.}\ }\textbf {\bibinfo {volume} {65}},\ \bibinfo
  {pages} {2442} (\bibinfo {year} {1990})}\BibitemShut {NoStop}%
\bibitem [{\citenamefont {Ketzmerick}(1996)}]{ketzmerick}%
  \BibitemOpen
  \bibfield  {author} {\bibinfo {author} {\bibfnamefont {R.}~\bibnamefont
  {Ketzmerick}},\ }\bibfield  {title} {\bibinfo {title} {Fractal conductance
  fluctuations in generic chaotic cavities},\ }\href
  {https://doi.org/10.1103/PhysRevB.54.10841} {\bibfield  {journal} {\bibinfo
  {journal} {Phys. Rev. B}\ }\textbf {\bibinfo {volume} {54}},\ \bibinfo
  {pages} {10841} (\bibinfo {year} {1996})}\BibitemShut {NoStop}%
\bibitem [{\citenamefont {Amin}\ \emph {et~al.}(2018)\citenamefont {Amin},
  \citenamefont {Ray}, \citenamefont {Pal}, \citenamefont {Pandit},\ and\
  \citenamefont {Bid}}]{Amin2018}%
  \BibitemOpen
  \bibfield  {author} {\bibinfo {author} {\bibfnamefont {K.~R.}\ \bibnamefont
  {Amin}}, \bibinfo {author} {\bibfnamefont {S.~S.}\ \bibnamefont {Ray}},
  \bibinfo {author} {\bibfnamefont {N.}~\bibnamefont {Pal}}, \bibinfo {author}
  {\bibfnamefont {R.}~\bibnamefont {Pandit}},\ and\ \bibinfo {author}
  {\bibfnamefont {A.}~\bibnamefont {Bid}},\ }\bibfield  {title} {\bibinfo
  {title} {Exotic multifractal conductance fluctuations in graphene},\ }\href
  {https://doi.org/10.1038/s42005-017-0001-4} {\bibfield  {journal} {\bibinfo
  {journal} {Communications Physics}\ }\textbf {\bibinfo {volume} {1}},\
  \bibinfo {pages} {1} (\bibinfo {year} {2018})}\BibitemShut {NoStop}%
\bibitem [{\citenamefont {M\"uller}\ \emph {et~al.}(2004)\citenamefont
  {M\"uller}, \citenamefont {Heusler}, \citenamefont {Braun}, \citenamefont
  {Haake},\ and\ \citenamefont {Altland}}]{PhysRevLett.93.014103}%
  \BibitemOpen
  \bibfield  {author} {\bibinfo {author} {\bibfnamefont {S.}~\bibnamefont
  {M\"uller}}, \bibinfo {author} {\bibfnamefont {S.}~\bibnamefont {Heusler}},
  \bibinfo {author} {\bibfnamefont {P.}~\bibnamefont {Braun}}, \bibinfo
  {author} {\bibfnamefont {F.}~\bibnamefont {Haake}},\ and\ \bibinfo {author}
  {\bibfnamefont {A.}~\bibnamefont {Altland}},\ }\bibfield  {title} {\bibinfo
  {title} {Semiclassical foundation of universality in quantum chaos},\ }\href
  {https://doi.org/10.1103/PhysRevLett.93.014103} {\bibfield  {journal}
  {\bibinfo  {journal} {Phys. Rev. Lett.}\ }\textbf {\bibinfo {volume} {93}},\
  \bibinfo {pages} {014103} (\bibinfo {year} {2004})}\BibitemShut {NoStop}%
\bibitem [{\citenamefont {Novaes}(2013)}]{Novaes_2013}%
  \BibitemOpen
  \bibfield  {author} {\bibinfo {author} {\bibfnamefont {M.}~\bibnamefont
  {Novaes}},\ }\bibfield  {title} {\bibinfo {title} {Combinatorial problems in
  the semiclassical approach to quantum chaotic transport},\ }\href
  {https://doi.org/10.1088/1751-8113/46/9/095101} {\bibfield  {journal}
  {\bibinfo  {journal} {Journal of Physics A: Mathematical and Theoretical}\
  }\textbf {\bibinfo {volume} {46}},\ \bibinfo {pages} {095101} (\bibinfo
  {year} {2013})}\BibitemShut {NoStop}%
\bibitem [{\citenamefont {Richter}\ and\ \citenamefont
  {Sieber}(2002)}]{PhysRevLett.89.206801}%
  \BibitemOpen
  \bibfield  {author} {\bibinfo {author} {\bibfnamefont {K.}~\bibnamefont
  {Richter}}\ and\ \bibinfo {author} {\bibfnamefont {M.}~\bibnamefont
  {Sieber}},\ }\bibfield  {title} {\bibinfo {title} {Semiclassical theory of
  chaotic quantum transport},\ }\href
  {https://doi.org/10.1103/PhysRevLett.89.206801} {\bibfield  {journal}
  {\bibinfo  {journal} {Phys. Rev. Lett.}\ }\textbf {\bibinfo {volume} {89}},\
  \bibinfo {pages} {206801} (\bibinfo {year} {2002})}\BibitemShut {NoStop}%
\bibitem [{\citenamefont {Berkolaiko}\ and\ \citenamefont
  {Kuipers}(2012)}]{PhysRevE.85.045201}%
  \BibitemOpen
  \bibfield  {author} {\bibinfo {author} {\bibfnamefont {G.}~\bibnamefont
  {Berkolaiko}}\ and\ \bibinfo {author} {\bibfnamefont {J.}~\bibnamefont
  {Kuipers}},\ }\bibfield  {title} {\bibinfo {title} {Universality in chaotic
  quantum transport: The concordance between random-matrix and semiclassical
  theories},\ }\href {https://doi.org/10.1103/PhysRevE.85.045201} {\bibfield
  {journal} {\bibinfo  {journal} {Phys. Rev. E}\ }\textbf {\bibinfo {volume}
  {85}},\ \bibinfo {pages} {045201} (\bibinfo {year} {2012})}\BibitemShut
  {NoStop}%
\bibitem [{\citenamefont {Kuipers}\ and\ \citenamefont
  {Richter}(2013)}]{Kuipers_2013}%
  \BibitemOpen
  \bibfield  {author} {\bibinfo {author} {\bibfnamefont {J.}~\bibnamefont
  {Kuipers}}\ and\ \bibinfo {author} {\bibfnamefont {K.}~\bibnamefont
  {Richter}},\ }\bibfield  {title} {\bibinfo {title} {Transport moments and
  andreev billiards with tunnel barriers},\ }\href
  {https://doi.org/10.1088/1751-8113/46/5/055101} {\bibfield  {journal}
  {\bibinfo  {journal} {Journal of Physics A: Mathematical and Theoretical}\
  }\textbf {\bibinfo {volume} {46}},\ \bibinfo {pages} {055101} (\bibinfo
  {year} {2013})}\BibitemShut {NoStop}%
\bibitem [{\citenamefont {Groth}\ \emph {et~al.}(2014)\citenamefont {Groth},
  \citenamefont {Wimmer}, \citenamefont {Akhmerov},\ and\ \citenamefont
  {Waintal}}]{kwant}%
  \BibitemOpen
  \bibfield  {author} {\bibinfo {author} {\bibfnamefont {C.~W.}\ \bibnamefont
  {Groth}}, \bibinfo {author} {\bibfnamefont {M.}~\bibnamefont {Wimmer}},
  \bibinfo {author} {\bibfnamefont {A.~R.}\ \bibnamefont {Akhmerov}},\ and\
  \bibinfo {author} {\bibfnamefont {X.}~\bibnamefont {Waintal}},\ }\bibfield
  {title} {\bibinfo {title} {Kwant: a software package for quantum transport},\
  }\href {https://doi.org/10.1088/1367-2630/16/6/063065} {\bibfield  {journal}
  {\bibinfo  {journal} {New Journal of Physics}\ }\textbf {\bibinfo {volume}
  {16}},\ \bibinfo {pages} {063065} (\bibinfo {year} {2014})}\BibitemShut
  {NoStop}%
\bibitem [{\citenamefont {Kantelhardt}\ \emph {et~al.}(2002)\citenamefont
  {Kantelhardt}, \citenamefont {Zschiegner}, \citenamefont {Koscielny-Bunde},
  \citenamefont {Havlin}, \citenamefont {Bunde},\ and\ \citenamefont
  {Stanley}}]{mfdfa}%
  \BibitemOpen
  \bibfield  {author} {\bibinfo {author} {\bibfnamefont {J.~W.}\ \bibnamefont
  {Kantelhardt}}, \bibinfo {author} {\bibfnamefont {S.~A.}\ \bibnamefont
  {Zschiegner}}, \bibinfo {author} {\bibfnamefont {E.}~\bibnamefont
  {Koscielny-Bunde}}, \bibinfo {author} {\bibfnamefont {S.}~\bibnamefont
  {Havlin}}, \bibinfo {author} {\bibfnamefont {A.}~\bibnamefont {Bunde}},\ and\
  \bibinfo {author} {\bibfnamefont {H.}~\bibnamefont {Stanley}},\ }\bibfield
  {title} {\bibinfo {title} {Multifractal detrended fluctuation analysis of
  nonstationary time series},\ }\href
  {https://doi.org/https://doi.org/10.1016/S0378-4371(02)01383-3} {\bibfield
  {journal} {\bibinfo  {journal} {Physica A: Statistical Mechanics and its
  Applications}\ }\textbf {\bibinfo {volume} {316}},\ \bibinfo {pages} {87}
  (\bibinfo {year} {2002})}\BibitemShut {NoStop}%
\bibitem [{\citenamefont {Mandelbrot}(1982)}]{Mandelbrot}%
  \BibitemOpen
  \bibfield  {author} {\bibinfo {author} {\bibfnamefont {B.~B.}\ \bibnamefont
  {Mandelbrot}},\ }\href {https://cds.cern.ch/record/98509} {\emph {\bibinfo
  {title} {{The fractal geometry of nature}}}}\ (\bibinfo  {publisher}
  {Freeman},\ \bibinfo {address} {San Francisco, CA},\ \bibinfo {year}
  {1982})\BibitemShut {NoStop}%
\bibitem [{\citenamefont {Suyari}(2006)}]{q-statistics}%
  \BibitemOpen
  \bibfield  {author} {\bibinfo {author} {\bibfnamefont {H.}~\bibnamefont
  {Suyari}},\ }\bibfield  {title} {\bibinfo {title} {Mathematical structures
  derived from the q-multinomial coefficient in tsallis statistics},\ }\href
  {https://doi.org/https://doi.org/10.1016/j.physa.2005.12.061} {\bibfield
  {journal} {\bibinfo  {journal} {Physica A: Statistical Mechanics and its
  Applications}\ }\textbf {\bibinfo {volume} {368}},\ \bibinfo {pages} {63}
  (\bibinfo {year} {2006})}\BibitemShut {NoStop}%
\bibitem [{\citenamefont {Frahm}\ and\ \citenamefont
  {Pichard}(1995)}]{frahm-pichard1}%
  \BibitemOpen
  \bibfield  {author} {\bibinfo {author} {\bibfnamefont {K.}~\bibnamefont
  {Frahm}}\ and\ \bibinfo {author} {\bibfnamefont {J.-L.}\ \bibnamefont
  {Pichard}},\ }\bibfield  {title} {\bibinfo {title} {Magnetoconductance of
  ballistic chaotic quantum dots: A brownian motion approach for the
  s-matrix},\ }\href {https://doi.org/10.1051/jp1:1995171} {\bibfield
  {journal} {\bibinfo  {journal} {Journal de Physique I}\ }\textbf {\bibinfo
  {volume} {5}},\ \bibinfo {pages} {847} (\bibinfo {year} {1995})}\BibitemShut
  {NoStop}%
\bibitem [{\citenamefont {Evers}\ and\ \citenamefont
  {Mirlin}(2008)}]{anderson-transitions}%
  \BibitemOpen
  \bibfield  {author} {\bibinfo {author} {\bibfnamefont {F.}~\bibnamefont
  {Evers}}\ and\ \bibinfo {author} {\bibfnamefont {A.~D.}\ \bibnamefont
  {Mirlin}},\ }\bibfield  {title} {\bibinfo {title} {Anderson transitions},\
  }\href {https://doi.org/10.1103/RevModPhys.80.1355} {\bibfield  {journal}
  {\bibinfo  {journal} {Rev. Mod. Phys.}\ }\textbf {\bibinfo {volume} {80}},\
  \bibinfo {pages} {1355} (\bibinfo {year} {2008})}\BibitemShut {NoStop}%
\bibitem [{\citenamefont {Salazar}\ and\ \citenamefont
  {Vasconcelos}(2010)}]{salazar-vasconcelos}%
  \BibitemOpen
  \bibfield  {author} {\bibinfo {author} {\bibfnamefont {D.~S.~P.}\
  \bibnamefont {Salazar}}\ and\ \bibinfo {author} {\bibfnamefont {G.~L.}\
  \bibnamefont {Vasconcelos}},\ }\bibfield  {title} {\bibinfo {title}
  {Stochastic dynamical model of intermittency in fully developed turbulence},\
  }\href {https://doi.org/10.1103/PhysRevE.82.047301} {\bibfield  {journal}
  {\bibinfo  {journal} {Phys. Rev. E}\ }\textbf {\bibinfo {volume} {82}},\
  \bibinfo {pages} {047301} (\bibinfo {year} {2010})}\BibitemShut {NoStop}%
\bibitem [{\citenamefont {Mac\^edo}\ \emph {et~al.}(2017)\citenamefont
  {Mac\^edo}, \citenamefont {Gonz\'alez}, \citenamefont {Salazar},\ and\
  \citenamefont {Vasconcelos}}]{pre-H}%
  \BibitemOpen
  \bibfield  {author} {\bibinfo {author} {\bibfnamefont {A.~M.~S.}\
  \bibnamefont {Mac\^edo}}, \bibinfo {author} {\bibfnamefont {I.~R.~R.}\
  \bibnamefont {Gonz\'alez}}, \bibinfo {author} {\bibfnamefont {D.~S.~P.}\
  \bibnamefont {Salazar}},\ and\ \bibinfo {author} {\bibfnamefont {G.~L.}\
  \bibnamefont {Vasconcelos}},\ }\bibfield  {title} {\bibinfo {title}
  {Universality classes of fluctuation dynamics in hierarchical complex
  systems},\ }\href {https://doi.org/10.1103/PhysRevE.95.032315} {\bibfield
  {journal} {\bibinfo  {journal} {Phys. Rev. E}\ }\textbf {\bibinfo {volume}
  {95}},\ \bibinfo {pages} {032315} (\bibinfo {year} {2017})}\BibitemShut
  {NoStop}%
\end{thebibliography}%

\end{document}